\documentclass[aps,prb,twocolumn,groupaddress,longbibliography]{revtex4-2}
\usepackage{epsfig,amsopn}
\usepackage{graphicx}
\usepackage{color}
\usepackage{amsmath,amssymb}
\usepackage{enumerate}

\newcommand\bea{\begin{eqnarray}}
\newcommand\eea{\end{eqnarray}}
\newcommand\beq{\begin{equation}}
\newcommand\eeq{\end{equation}}

\def\nn{\nonumber}
\def\f{\frac}
\def\mL{\mathcal{L}}
\def\mR{\mathcal{R}}
\def\al{\alpha}

\def\ga{\gamma}

\def\si{\sigma}
\def\Do{\partial}
\def\De{\Delta}

\def\ra{\rangle}
\def\ua{\uparrow}
\def\da{\downarrow}

\def\be{\beta}

\def\th{\theta}

\DeclareUnicodeCharacter{0308}{\"}

\begin{document}
  \title{Crossed Andreev reflection in collinear $p$-wave magnet/triplet superconductor junctions}
  \author{ Abhiram Soori}
  \email{abhirams@uohyd.ac.in}
  \affiliation{School of Physics, University of Hyderabad, Prof. C. R. Rao Road, Gachibowli, Hyderabad-500046, India}
\begin{abstract} 
Crossed Andreev reflection (CAR) is a fundamental quantum transport phenomenon that holds significant implications for spintronics and superconducting devices. However, its experimental detection and enhancement remain challenging. Recently, magnetic materials exhibiting $p$-wave magnetic ordering, distinct from conventional spin-orbit coupling, referred to as $p$-wave magnets, have attracted considerable interest. In this work, we propose a junction consisting of $p$-wave magnets and a triplet superconductor as a promising platform to enhance CAR. The setup features a triplet superconductor sandwiched between two collinear $p$-wave magnets rotated by $180^\circ$ relative to each other, allowing for precise control over transport processes. We demonstrate that CAR can dominate over electron tunneling (ET) within specific parameter regimes, such as the orientation angle of the $p$-wave magnets and their chemical potential. Enhanced CAR occurs when the constant energy contours of the two spins in the $p$-wave magnets are well-separated. Furthermore, the conductivities display Fabry-Pérot-type oscillations due to interference effects, with CAR diminishing as the length of the superconductor exceeds the decay length of the wavefunctions. These findings underscore the potential of collinear $p$-wave magnet-superconductor junctions as a robust platform for the experimental investigation and enhancement of CAR.
\end{abstract}
\maketitle
\section{Introduction}
The concept of altermagnetism has garnered substantial attention from both theorists and experimentalists in recent years~\cite{smejkal22a,smejkal22b,smejkal22c,jung2024rev}. In magnetic materials, analogues of various superconducting phases can be identified by mapping the particle-hole sectors of a superconductor onto the two spin sectors. Altermagnets can be viewed as magnetic analogs of $d$-wave superconductors. In contrast, the magnetic counterparts of $p$-wave superconductors are found in spin-orbit-coupled materials, while the magnetic equivalent of the anisotropic triplet pairing observed in liquid helium-3 corresponds to $p$-wave magnets~\cite{jung24pwave,sukhachov24fr,hedayati25,ezawa24,sukhachov24co}. A defining characteristic of both altermagnets and $p$-wave magnets is their spin-split band structure, which significantly influences their transport properties.

\begin{figure}[htb]
\includegraphics[width=8cm]{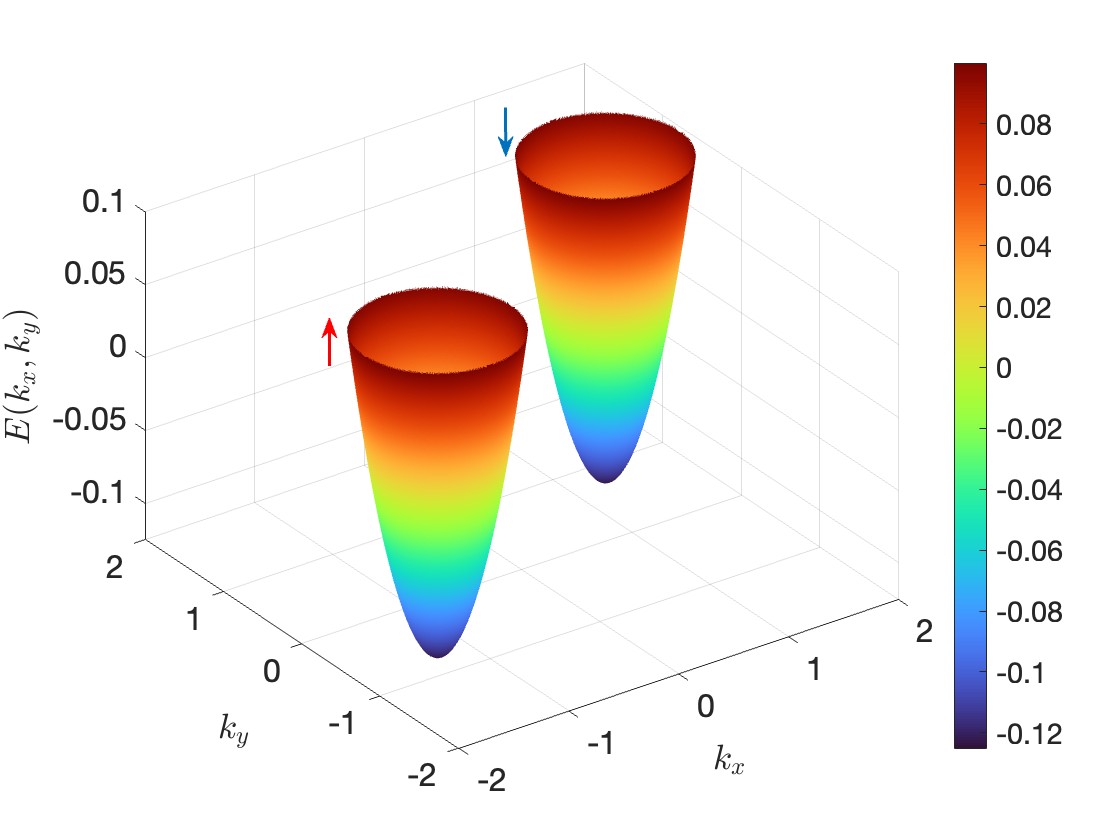}
\caption{Bandstructure of a $p$-wave magnet. The bands for the two spins are well separated. 
}\label{fig:band}
\end{figure}

\noindent Unlike $d$-wave altermagnets, $p$-wave magnets maintain time-reversal symmetry. Although $p$-wave magnets exhibit qualitative similarities to spin-orbit-coupled systems, their band structure is highly anisotropic at $\vec k = 0$ (see Fig.~\ref{fig:band}). In two dimensions, spin-orbit coupling introduces a term in the Hamiltonian that is proportional to $(\sigma_x k_y - \sigma_y k_x)$, while $p$-wave magnets are characterized by the term $\sigma_z k_x$, resulting in an anisotropic Fermi surface. Both $d$-wave altermagnets and $p$-wave magnets exhibit zero net spin polarization. However, a voltage-biased junction formed between $d$-wave altermagnets and metals has been shown to carry a spin current~\cite{das2023}, while a similarly biased junction between $p$-wave magnets and metals generates a transverse spin current~\cite{hedayati25}.

\noindent Andreev reflection (AR) is a fundamental process that occurs at the interface between a superconductor and another material, where an incoming electron is reflected as a hole, facilitating the formation of a Cooper pair within the superconductor. In the context of two-dimensional altermagnet-singlet superconductor junctions, maximal AR is observed for specific orientations of the altermagnet, attributable to its spin-split band structure~\cite{sun23ar,papaj}.

\noindent Crossed Andreev reflection (CAR) expands upon conventional Andreev reflection by facilitating electron-to-hole conversion across spatially separated regions, such as different $p$-wave magnets connected via a superconductor. Although CAR has been extensively investigated in junctions involving normal metals, ferromagnets, altermagnets, and superconductors~\cite{beckmann04,yamashita03,soori17,zhang2019,nehra19,soori19,Jakobsen21,soori22car,Zhao,das2024}, its experimental realization remains challenging. A significant hurdle is that electron tunneling (ET) through the superconductor often overshadows CAR, complicating its detection. CAR is essential not only for probing superconducting pairing symmetry~\cite{benjamin2006,crepin2015}  but also for its role in Cooper pair splitting~\cite{recher2001,Das2012,schin12}, which allows for the generation of spatially separated and entangled electrons. Understanding the pairing symmetry in triplet superconductors is particularly critical, as evidenced by the increasing number of candidate materials, such as Sr$_2$RuO$_4$, NiBi, and UTe$_2$, which demonstrate triplet superconductivity~\cite{Ishida1998,Luke1998,he2022,wu2024ute2}.

\noindent A commonly employed strategy to enhance crossed Andreev reflection (CAR) involves the use of ferromagnets, where spin filtering reduces electron tunneling (ET) and Andreev reflection (AR)~\cite{beckmann04}. However, this approach generally necessitates external magnetic fields to align the spins, which adds to the experimental complexity. In contrast, $p$-wave magnets naturally possess spin-split band structures due to their intrinsic magnetic ordering, thereby eliminating the need for external fields. We demonstrate that in a junction where a triplet superconductor is situated between two $p$-wave magnets with their orientations rotated by $180^\circ$ relative to each other, CAR can be selectively enhanced while completely suppressing both ET and AR. This configuration also offers a robust platform for probing triplet superconductivity.

\noindent In this work, we propose a robust mechanism for enhancing crossed Andreev reflection (CAR) in hybrid structures by leveraging the intrinsic spin-split band structure of $p$-wave magnets in contact with a triplet superconductor. Using Landauer-Büttiker scattering theory, we analyze electron transport in a junction where the two $p$-wave magnets are rotated by $180^\circ$ relative to each other. We demonstrate that for specific orientations of the magnets, CAR becomes the dominant transport process, while electron tunneling (ET) and Andreev reflection (AR) are completely suppressed. This suppression of ET and AR arises due to transverse momentum matching: within the selected range of orientations, no electron states with the same spin are available on both sides of the superconductor, preventing ET, while the absence of same-spin electron and hole states on the incident side blocks AR. Unlike conventional methods that rely on ferromagnets, where suppression of ET requires precise spin filtering and external magnetic fields, our approach achieves this suppression naturally due to the inherent properties of $p$-wave magnets. These findings position $p$-wave magnet-superconductor junctions as an ideal platform for experimentally realizing and controlling CAR, paving the way for applications in superconducting spintronics.

\begin{figure}
\includegraphics[width=8cm]{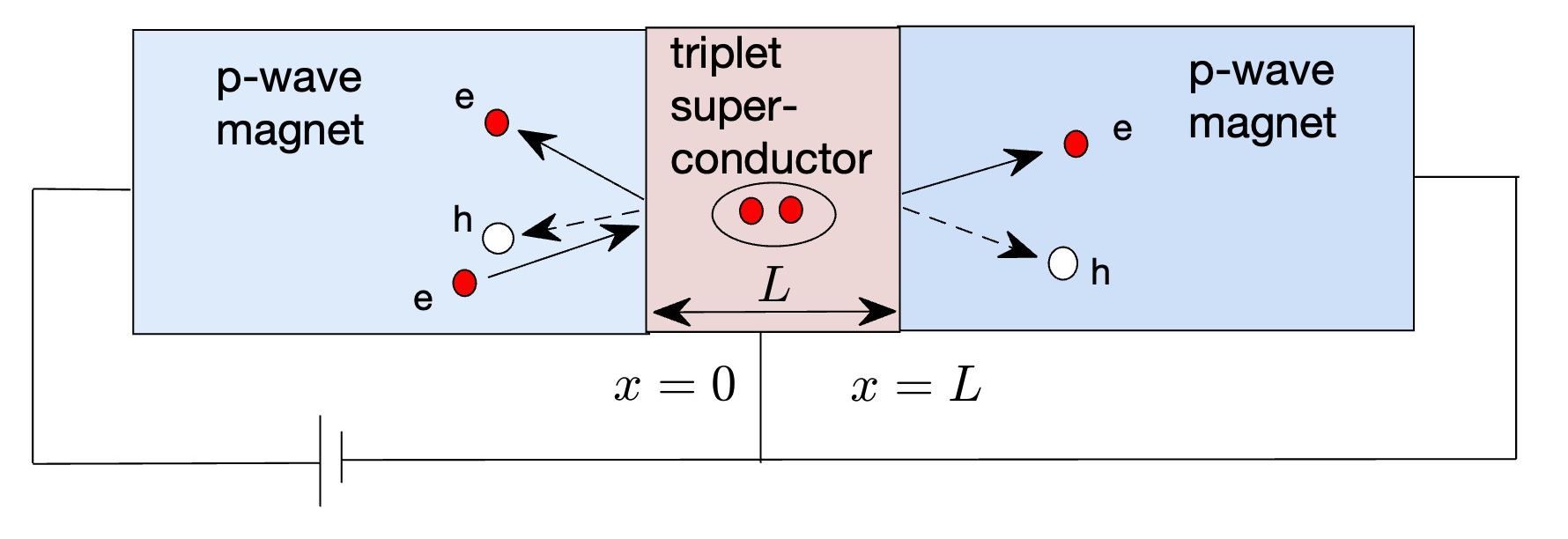}
\caption{Setup: Two $p$-wave magnets are connected on the two sides of a triplet superconductor. Orientation of the two $p$-wave magnets are rotated by 180$^{\circ}$ with respect to one another. }\label{fig:schem}
\end{figure}

\section{ System and Key Result} 
\begin{figure}
\includegraphics[width=4cm]{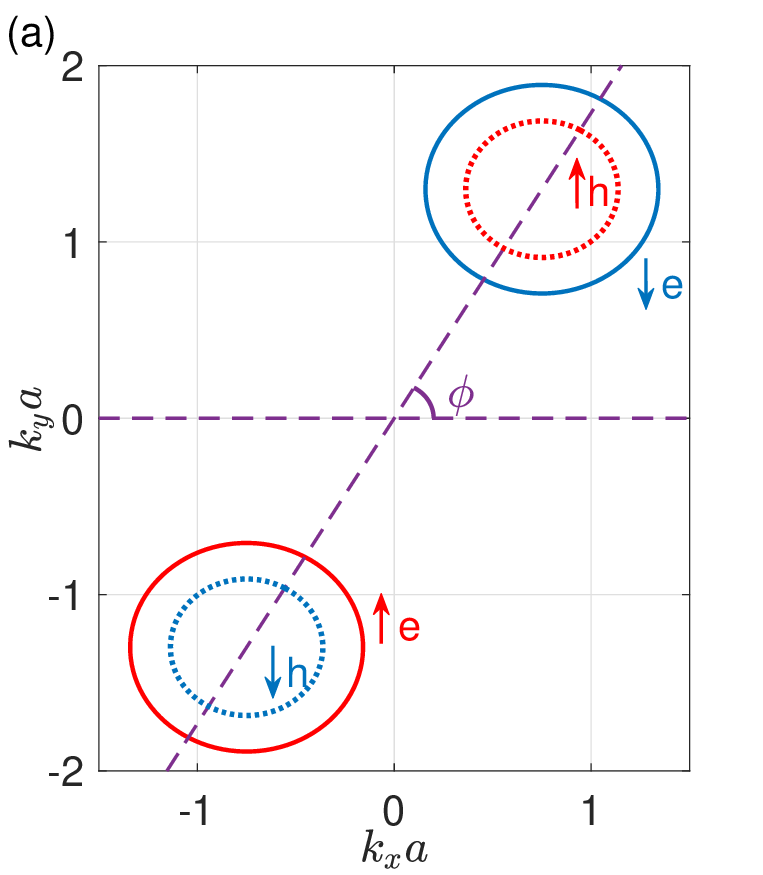}
\includegraphics[width=4cm]{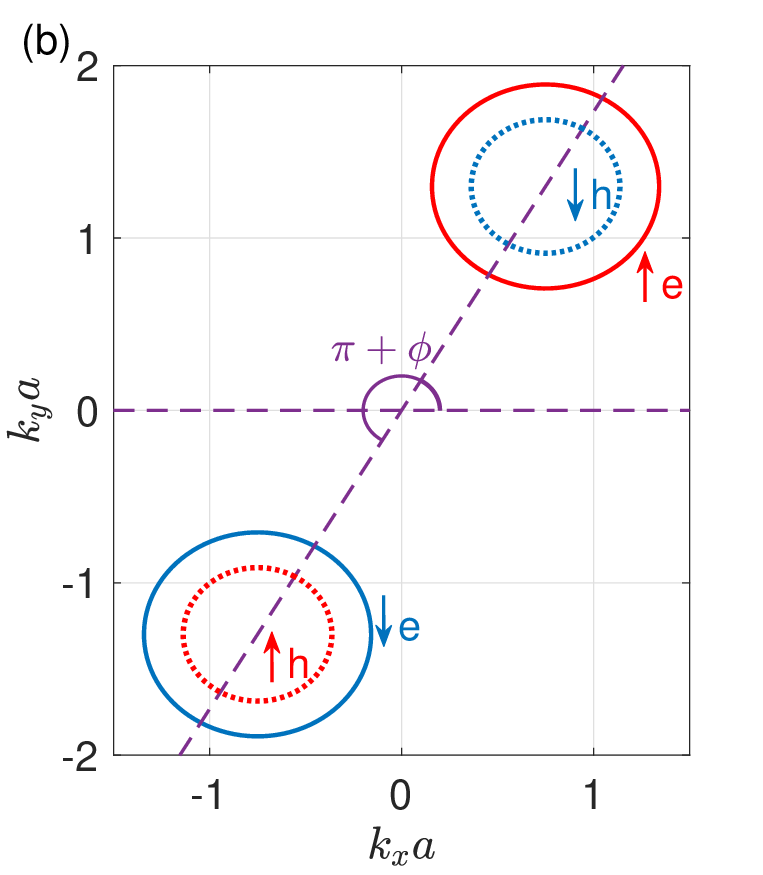}
\caption{Constant energy contours of the left (a) and right (b) $p$-wave magnets. Note that $k_y$ of electrons and holes of the same spin on either sides match in a certain range, giving scope for CAR. Parameters: $\mu=-\mu_s$, $\al=1.5\hbar m/a$, $\phi=\pi/3$ and $E=0.05\mu_s$.}\label{fig:cec}
\end{figure}

We consider a junction where $p$-wave magnets are attached on either side of a triplet superconductor, with the $p$-wave magnet on the right rotated by $180^{\circ}$ relative to the one on the left. A schematic diagram of the setup is shown in Fig.~\ref{fig:schem}.  Fig.~\ref{fig:cec}(a) and Fig.~\ref{fig:cec}(b)  illustrate the constant energy contours of the dispersion relations for the two $p$-wave magnets. These figures reveal that the transverse wavenumber $k_y$ in a range matches across the junction for electrons on the left and holes on the right, both having same spin. Due to this,  incident electrons from the left $p$-wave magnet scatter into hole states on the right, provided equal spin electron-to-hole conversion occurs in the central superconducting region. This conversion is facilitated by the triplet pairing in the superconductor.

\section{Details of calculation}
\subsection{ Hamiltonian } The Hamiltonian for the system is given by  
\bea 
H &=& \begin{cases} 
\Big[\f{-\hbar^2 \vec \nabla^2}{2m}-\mu\Big]\tau_z\si_0-i \al\hbar(\hat n_{\phi} \cdot \vec \nabla)\tau_0\si_z, \\~~~~~~~~~~~~~~~~~~~~~~~~{\rm for}~~x<0,  \\ 
 \Big[\f{-\hbar^2 \vec \nabla^2}{2m}-\mu_s\Big]\tau_z\si_0 - \f{i\De\Do_x}{k_F}\tau_x\si_{\theta}, \\~~~~~~~~~~~~~~~~~~~~~~~~{\rm for}~~0<x<L, \\ 
  \Big[\f{-\hbar^2 \vec \nabla^2}{2m}-\mu\Big]\tau_z\si_0-i \al\hbar(\hat n_{\pi+\phi} \cdot \vec \nabla)\tau_0\si_z, \\~~~~~~~~~~~~~~~~~~~~~~~~{\rm for}~~x>L, 
  \end{cases}
 \label{eq:ham}
\eea
where $m$ is the effective mass of electrons, $\mu$ ($\mu_s$) is the chemical potential in the $p$-wave magnet (superconductor), and $\al$ is the strength of the $p$-wave magnet term. The unit vector $\hat n_{\phi}=(\cos\phi\hat x+\sin\phi\hat y)$ makes an angle $\phi$ with $\hat x$. The Pauli matrices $\tau_j$ and $\si_j$ act on the particle-hole and spin sectors, respectively, with $j=0, x, y, z$. Here, $\vec \nabla=\hat x\Do_x+\hat y \Do_y$, $k_F=\sqrt{2m\mu_s/\hbar^2}$, and $\si_{\th}=\cos{\th}\si_z+\sin{\th}\si_x$, where $\th$ is the angle made by the spin direction of the equal-spin triplet pairing in the superconductor with $\hat z$, the easy axis of the $p$-wave magnet. The length of the central superconducting region is $L$.  The Hamiltonian acts on the four-spinor $\Psi = [\psi_{e,\ua}, \psi_{e,\da}, \psi_{h,\ua}, \psi_{h,\da}]^T$, where subscripts denote particle type and spin. The superconducting term in the Hamiltonian (in the region $0<x<L$) has been chosen to be of the form $\De k_x\tau_x\si_{\th}/k_F$, from a more general form: $\De (a_xk_x+a_yk_y)\tau_x\si_{\th}/k_F$, where $a_x, a_y$ are real numbers~\cite{soori13}. The choice of $\th$ decides the direction of spins in the equal-spin triplet pairing. For example, $\th=0$ corresponds to $(|\ua\ua\ra-|\da\da\ra)$, and $\th=\pi/2$ corresponds to $(|\ua\da\ra+|\da\ua\ra)$. 

A voltage bias $V$ is applied from the left $p$-wave magnet, with the superconductor and the right $p$-wave magnet grounded. The currents $I_{\mL}$ and $I_{\mR}$ in the left and right regions, and the differential conductivities $G_{\mL\mL}=dI_{\mL}/dV$ and $G_{\mR\mL}=dI_{\mR}/dV$ are calculated. 

\subsection{ Dispersion} The dispersion relation in the $p$-wave magnet is given by  
\bea 
E=\eta_{p}\Big[\f{\hbar^2(k_x^2+k_y^2)}{2m}-\mu\Big]+\ga \eta_{s}\al\hbar(\cos\phi k_x+\sin\phi k_y),  
\eea
where $p=e, h$ denotes electrons or holes, $s=\ua, \da$ denotes the spin, $\eta_{e/h}=\pm 1$, $\eta_{\ua/\da}=\pm 1$, and $\ga=1$~($-1$) for $x<0$ ($x>L$). The opposite signs of $\ga$ in $x<0$ and $x>L$ reflect the $180^\circ$ rotation of the $p$-wave magnets relative to one another. A typical dispersion relation is plotted in Fig.~\ref{fig:band}. The wavenumbers $k_x$ and $k_y$ for a given energy $E$ are parametrized as  
\bea 
k_x(p,E,\chi) &=& k_{0,p}\cos{\chi}-\ga\eta_s\eta_p\al m\cos{\phi}/\hbar, \nn \\
k_y(p,E,\chi) &=& k_{0,p}\sin{\chi}-\ga\eta_s\eta_p\al m\sin{\phi}/\hbar, \label{eq:kxky}
\eea
where $k_{0,p}=\sqrt{(\al m/\hbar)^2+2m(\mu+\eta_pE)/\hbar^2}$, and $\chi$ is the angle of the velocity vector relative to $\hat x$. 

The dispersion relation in the superconductor is given by  
\bea
E = \pm\sqrt{\Big[ \f{\hbar^2(k_x^2+k_y^2)}{2m}-\mu_s\Big]^2+\f{k_x^2}{k_F^2}\De^2}. \label{eq:dispsc}
\eea

\subsection{Boundary Conditions} To solve the scattering problem, boundary conditions are required in addition to the Hamiltonian in each region. The following boundary conditions ensure the conservation of probability current:  
\bea 
&&\Psi_P = c\Psi_S, \nn\\
&&c\Big[\f{\hbar}{m}\si_0\tau_z\Do_x\Psi_P+i\al\cos{\phi}\si_z\tau_0\Psi_P\Big] \nn \\ 
&&~~~~= \f{\hbar}{m} \si_0\tau_z\Do_x\Psi_S
+\f{\hbar \eta_{u} q}{m}\si_0\tau_z\Psi_S+ \f{i \De}{k_F}\tau_x\si_{\th}\Psi_S,~~~~\label{eq:bc}
\eea
where $c$ is a real, dimensionless parameter characterizing the junction transparency, and $q$ is a real parameter with dimensions of wavenumber, corresponding to the strength of a delta-function barrier near the junction on the superconducting side. Here, $\Psi_P$ and $\Psi_S$ are the wavefunctions on the $p$-wave magnet and superconductor sides of the junction, evaluated at $x=0$ and $x=L$. The parameter $\eta_u=1$ at $x=0$ and $\eta_u=-1$ at $x=L$. Physically, $c$ represents the hopping strength of the bond connecting the $p$-wave magnet to the superconductor in an equivalent lattice model~\cite{soori23scat} and we set $c=1$.  

\subsection{Scattering Eigenfunctions} Translational symmetry along $\hat y$ ensures that momentum along $\hat y$ is conserved. The eigenfunction of the Hamiltonian for an $s$-spin electron incident from the left $p$-wave magnet at energy $E$ and angle of incidence $\chi$ has the form $\Psi(x)e^{ik_yy}$, where $k_y=k_y(e,E,\chi)$ is obtained from Eq.~\eqref{eq:kxky} with $\ga=1$ and $-\pi/2<\chi<\pi/2$. The spatial part $\Psi(x)$ is expressed as:  
\bea
\Psi(x) &=& 
\begin{cases}
e^{ik^i_{x,e,s}x}|e,s\ra + \sum_{p,s'}r_{p,s',e,s}e^{ik^l_{x,p,s'}x}|p,s'\ra,  \\ 
~~~~~~~~~~~~~~~~~~~~~~~~~~~{\rm for}~~x<0, \\
\sum_{j',j,\si} s_{j',j,\si}e^{\si ik_{x,j}x} |\phi_{j',j,\si}\ra, \\
~~~~~~~~~~~~~~~~~~~~~~~~~~~{\rm for}~~0<x<L, \\
\sum_{p,s'} t_{p,s',e,s} e^{ik^r_{x,p,s'}(x-L)}|p,s'\ra, \\ 
~~~~~~~~~~~~~~~~~~~~~~~~~~~{\rm for}~~x>L,
\end{cases}
\eea
where $k^i_{x,e,s}=k_x(e,E,\chi)$ [see Eq.~\eqref{eq:kxky}] for $\ga=1$, and $k^l_{x,p,s'}$ ($k^r_{x,p,s'}$) are the $x$-components of the wavenumber for left-moving (right-moving) $p$-type particles with spin $s'$ in the regions $x<0$ ($x>L$), calculated using the respective dispersion relation with the same choice of $k_y$.  Here, when  wavenumber of a particle ($k^l_{x,p,s'}$ or $k^r_{x,p,s'}$) becomes complex, the one that gives exponentially decaying wavefunction away from the junction is chosen out of two possible solutions.  The spinors $|p,s\ra$ are given by $|e,\ua\ra=[1,0,0,0]^T$, $|e,\da\ra=[0,1,0,0]^T$, $|h,\ua\ra=[0,0,1,0]^T$, and $|h,\da\ra=[0,0,0,1]^T$. 

In the superconducting region, $k_x=\pm k_{x,j}$ ($j=1,2$) are the wavenumbers at energy $E$, obtained from the dispersion in Eq.~\eqref{eq:dispsc}. The corresponding eigenspinor $|\phi_{j',j,\si}\ra$ is associated with $k_x=\si k_{x,j}$, where $\si=\pm 1$, and is doubly degenerate, as captured by the index $j'=1,2$. The scattering coefficients $r_{p,s',e,s}$, $s_{j',j,\si}$, and $t_{p,s',e,s}$ are unknowns that are determined using the boundary conditions in Eq.~\eqref{eq:bc}.

\subsection{ Conductivities} Once the scattering coefficients are determined, local and nonlocal differential conductivities $G_{\mL\mL}$ and $G_{\mR\mL}$ can be calculated. The conductivities are given by
\bea 
G_{\mL\mL} &=& \f{e}{h^2}m\int_{-\pi/2}^{\pi/2}d\chi~ [J^c_{x,\ua,\mL}( \chi)+J^c_{x,\da,\mL}( \chi)] \nn \\ 
G_{\mR\mL} &=& \f{e}{h^2}m\int_{-\pi/2}^{\pi/2}d\chi~ [J^c_{x,\ua,\mR}( \chi)+J^c_{x,\da,\mR}( \chi)],  
\eea
where $J^c_{x,s,\mL/\mR}$ are the charge current densities carried in the region $x<0$ ($\mathcal{L}$) and $x>L$ ($\mathcal{R}$), when a spin-$s$ electron is incident from the left. The currents on the left  have the form  $J^c_{x,s,\mL}=e\hbar k_{0,e}\cos\chi/m+J^{c,r}_{x,s,\mL}$, where the first term corresponds to current due to incident spin-$s$ electron and the second term refers to the current due to all reflected electrons and holes. 
\bea 
J^{c,r}_{x,s,\mL} = e\sum_{p,s'}\Big[\f{\hbar k_{x,p,l,s'}}{m}+\eta_p\eta_s\al\cos\phi\Big]|r_{p,s',e,s}|^2\be_{\mL,p,s'}, \label{eq:JLr} 
\eea
where $\be_{\mL,p,s'}=1$ if $k_{x,p,l,s'}$ is purely real and zero otherwise. This is because, evanescent waves do not carry current in leads. In a similar way, the current on the right magnet is given by 
\bea 
J^{c}_{x,s,\mR} = e\sum_{p,s'}\Big[\f{\hbar k_{x,p,r,s'}}{m}-\eta_p\eta_s\al\cos\phi\Big]|t_{p,s',e,s}|^2\be_{\mR,p,s'}, \label{eq:JRt}
\eea
where $\be_{\mR,p,s'}=1$ if $k_{x,p,r,s'}$ is purely real and zero otherwise.

\begin{figure}[htb]
\includegraphics[width=6cm]{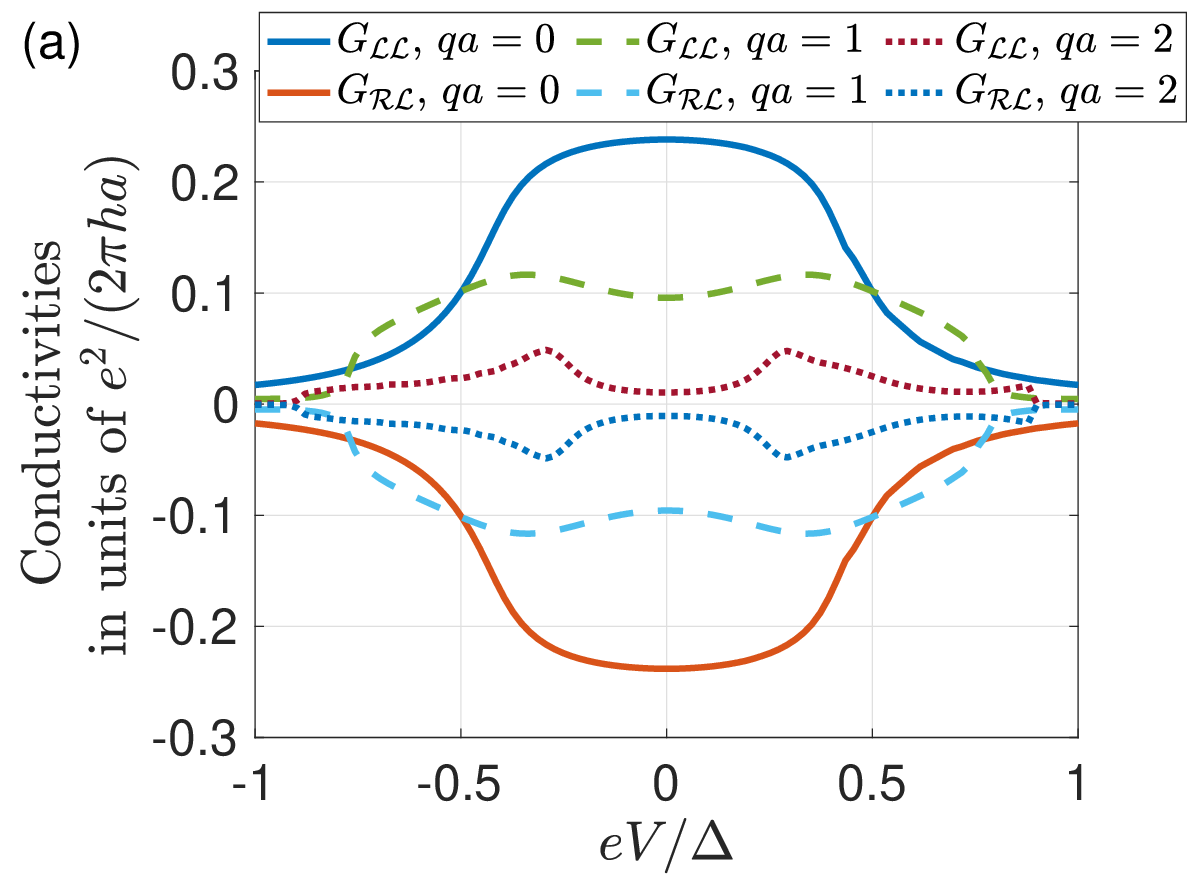}
\includegraphics[width=6cm]{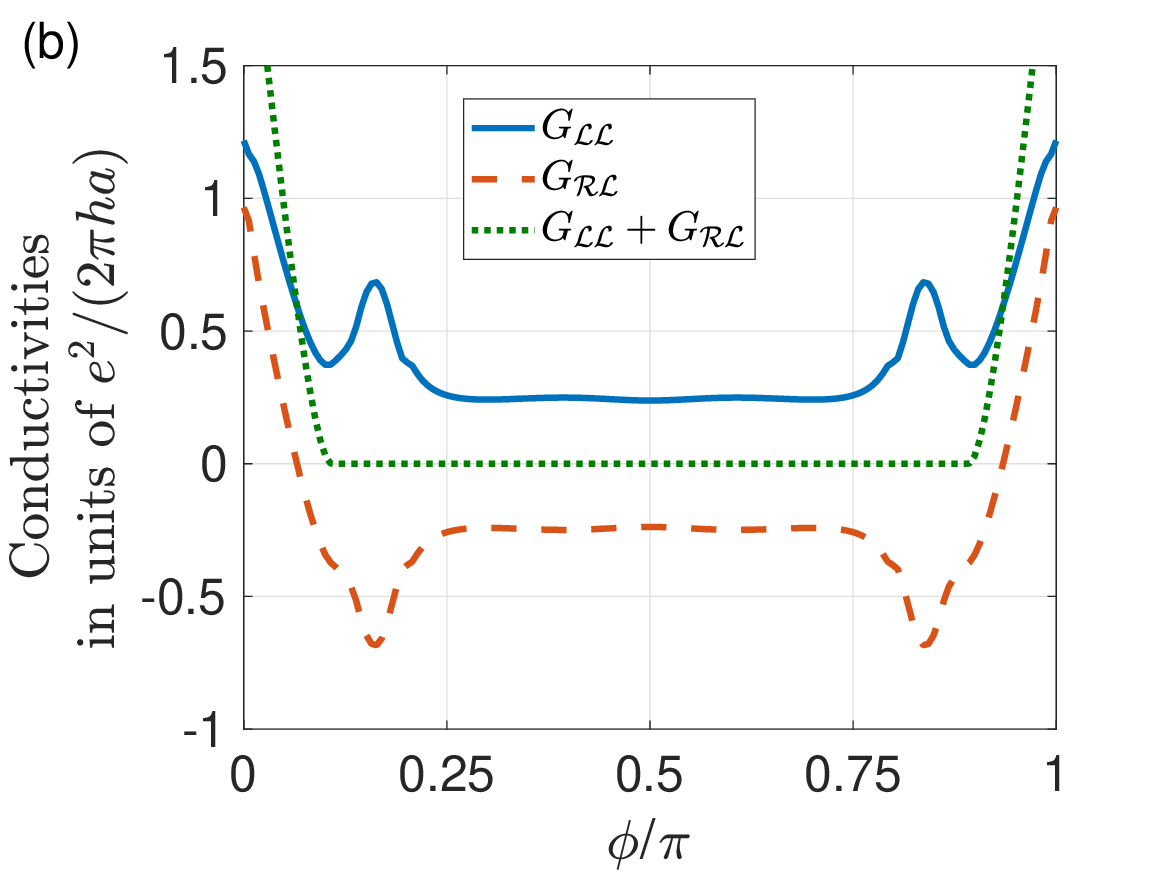}
\caption{Conductivities $G_{\mL\mL}$ and $G_{\mR\mL}$ versus: (a) bias $V$ for $\phi=\pi/2$, (b) orientation angle $\phi$ for $V=0$. In (a), different curves are for different values of $q$ - the barrier strength at the junction.  Parameters: $\mu=-\mu_s$, $\al=1.5\hbar /(ma)$, $\phi=\pi/2$, $\De=0.1\mu_s$, $\th=0$ and $L=5a$. }\label{fig:GvsVphi}
\end{figure}

\section{ Results} Parameters are expressed in units of $m$ and $\mu_s$, with $a = \hbar/\sqrt{m\mu_s}$ as the characteristic length scale. Setting $\mu=-\mu_s$, $\al=1.5\hbar/(ma)$, $\phi=\pi/2$, $\De=0.1\mu_s$, $\th=0$, and $L=5a$, the conductivities $G_{\mL\mL}$ and $G_{\mR\mL}$ are calculated and plotted in Fig.~\ref{fig:GvsVphi} as functions of (a) the bias $V$ and, (b) the orientation angle $\phi$. For $\th=0$, the pairing occurs between electrons with the same spin oriented along $\pm\hat z$. Under these conditions, an incident electron from the left can either reflect back as an electron or transmit to the right as a hole via CAR, resulting in $G_{\mL\mL}=-G_{\mR\mL}$. For this to happen, the constant energy contours for the two spins need to be well separated (as in Fig.~\ref{fig:cec}). 

The midgap Andreev bound states at the two ends of the superconductor hybridize, facilitating electron-to-hole conversion and enhancing CAR near zero bias.  In Fig.~\ref{fig:GvsVphi}(b), the conductivities are plotted as functions of $\phi$ with the bias fixed near zero. It is observed that in the range $0.11\pi<\phi<0.89\pi$, the local and nonlocal conductivities sum-up to zero, indicating absence of AR. Outside this range, same-spin electron-hole pairs with identical $k_y$ exist on the left, enabling AR. Additionally, same-spin electrons with the same $k_y$ are present on both sides of the superconductor, allowing ET. As a result, the two conductivities no longer sum to zero, and the nonlocal conductivity becomes positive near $\phi=0$, indicating that ET dominates over CAR.

\begin{figure}[htb]
\includegraphics[width=6cm]{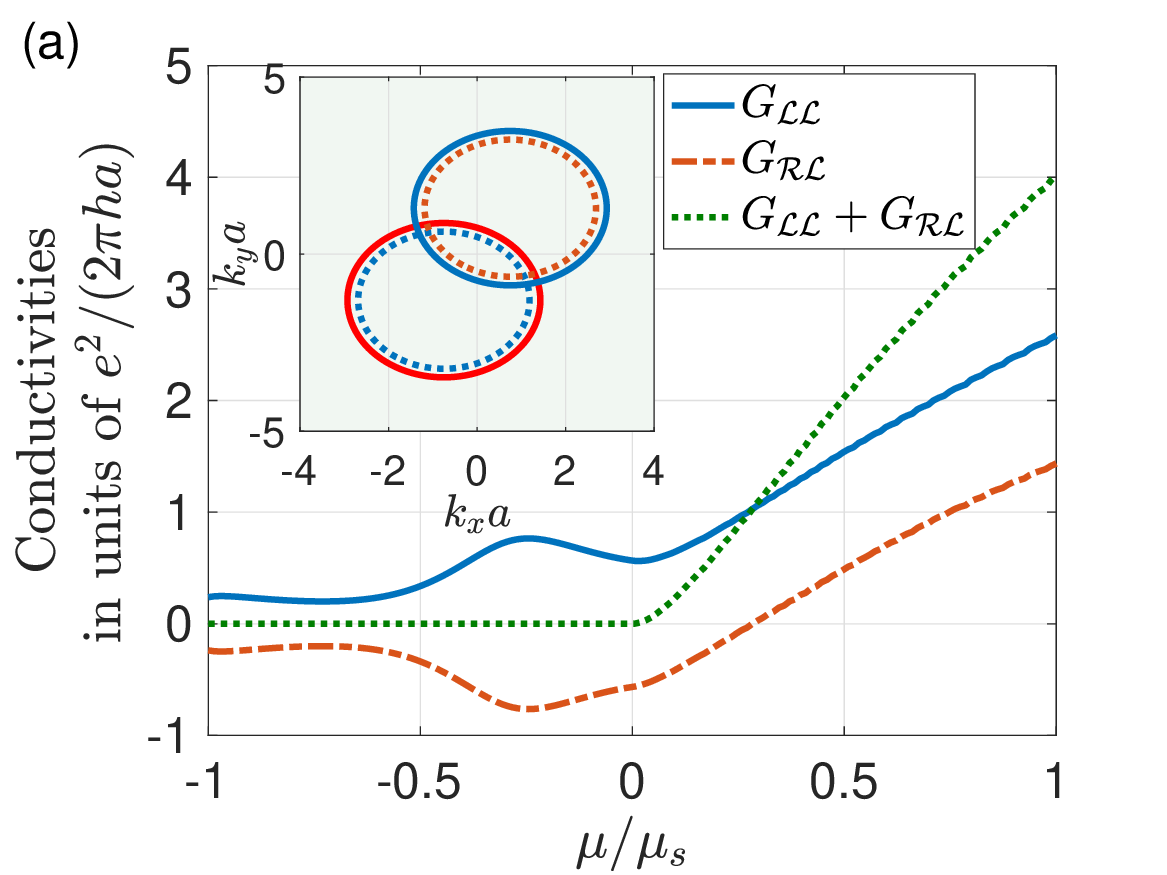}
\includegraphics[width=6cm]{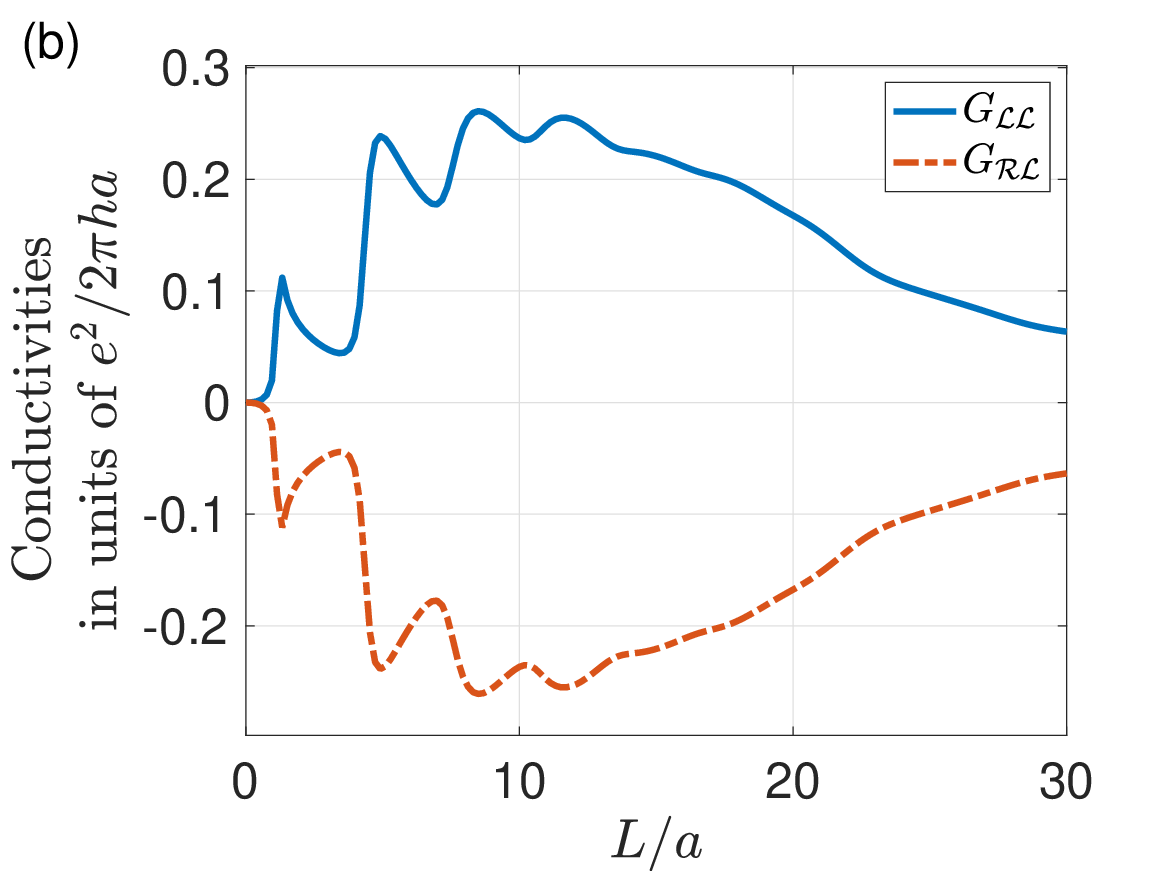}
\caption{Conductivities at zero bias versus: (a) $\mu$ - the chemical potential of the $p$-wave magnet for $q=0$ and $L=5$, (b) $L$ - the length of the superconductor for  $q=0$ and $\mu=-\mu_s$. Inset of (a): Constant energy contours for $\mu=\mu_s$, $\phi=\pi/3$, $E=0.25\mu_s$. Solid (dotted) line indicates electron (hole), and red (blue) color indicates spin-$\ua$~($\da$).  Other parameters same as in Fig.~\ref{fig:GvsVphi}. }\label{fig:GvsmuL}
\end{figure}

As a function of the chemical potential $\mu$ of the $p$-wave magnet, the conductivities exhibit the behavior shown in Fig.~\ref{fig:GvsmuL}(a). For $\mu_s > 0$, the two conductivities do not sum to zero. This is due to the crossing of constant energy contours for the two spins on each $p$-wave magnet, as illustrated in the inset of Fig.~\ref{fig:GvsmuL}(a), which allows for both AR and ET. For chemical potentials $\mu > 0.283\mu_s$, the sum of the conductivities becomes positive, indicating the dominance of ET over CAR.

The dependence of the conductivities on the length $L$ of the superconductor is also significant. Since CAR arises from electron-to-hole conversion within the superconductor, CAR initially increases as $L$ grows from zero. However, beyond a certain length, the exponentially decaying wavefunctions in the superconductor reduce the probability of the converted hole reaching the other $p$-wave magnet. The oscillatory behavior in the conductivities results from the real part of the wavenumbers in the superconductor, producing Fabry-P\'erot-type oscillations~\cite{soori17,sahu23}. For normal incidence, $k_{x1} = (1.4124 + 0.0707i)/a$, giving a decay length of $1/{\rm Im}[k_{x1}] = 14a$, where the conductivity peaks. The oscillation period is approximately $\pi/{\rm Re}[k_{x1}] = 2.22a$, consistent with the observed results.

\begin{figure}[htb]
\includegraphics[width=6cm]{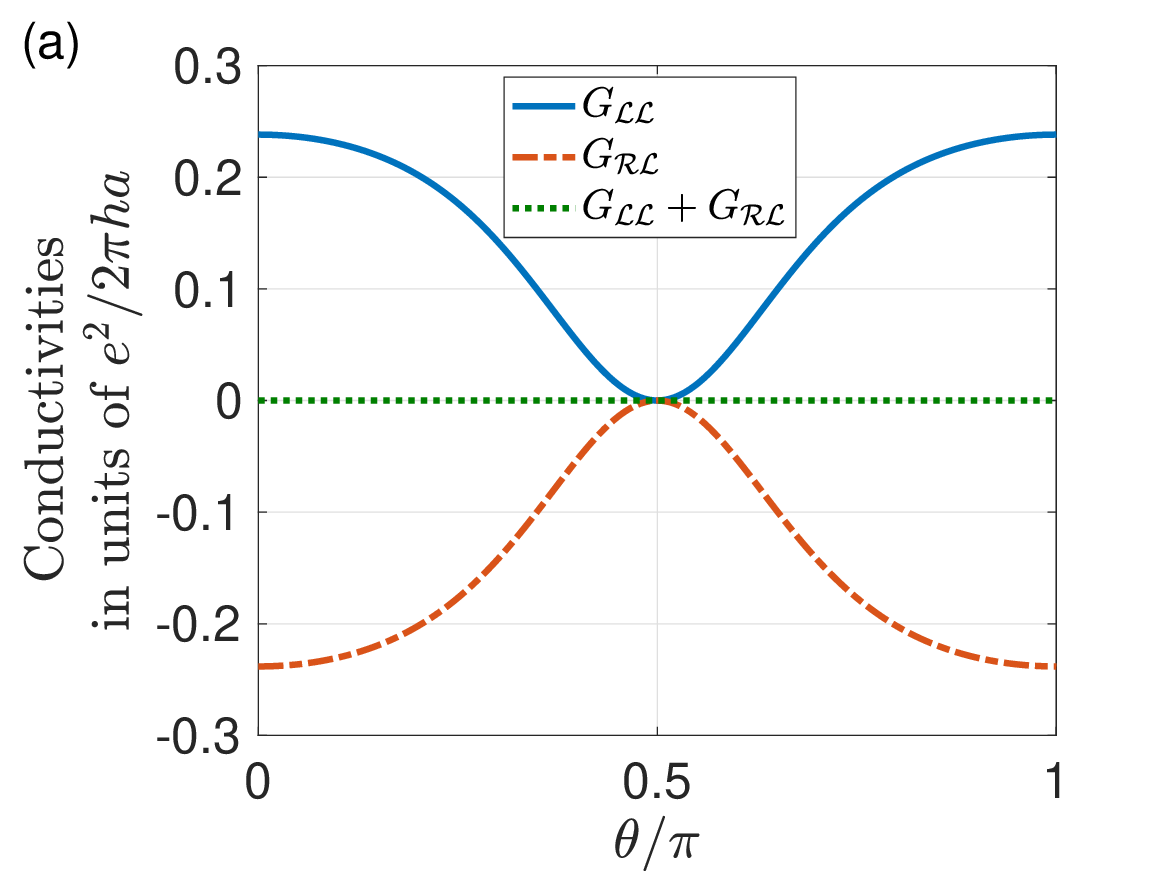}
\includegraphics[width=6cm]{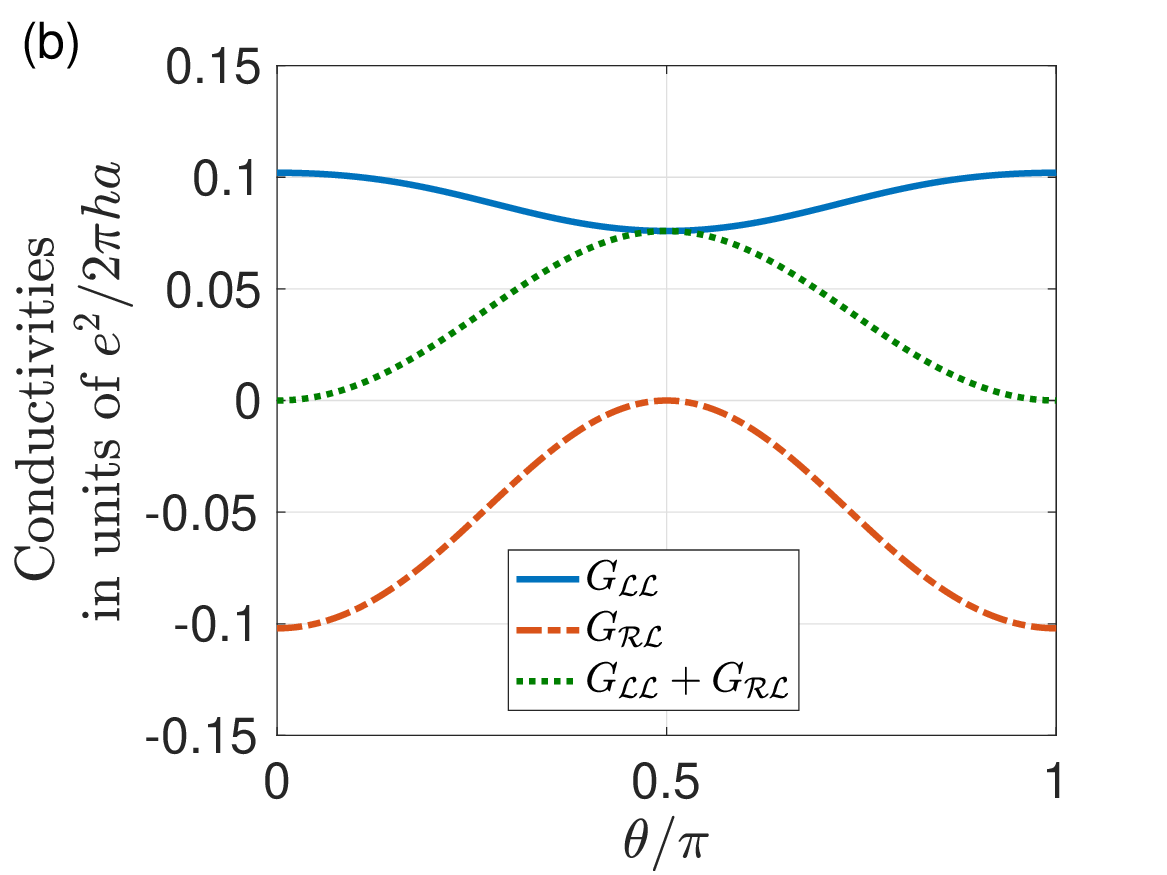}
\caption{Conductivities  versus the angle $\th$ between $\hat z$ and the direction of spins that make equal spin triplet pairing in superconductor for (a)  $V=0$ and  (b) $eV=0.5\De$. $q=0$   and other parameters same as in Fig.~\ref{fig:GvsVphi}. }\label{fig:Gvsth}
\end{figure}

Next, we examine the dependence of the conductivities on the angle $\th$ between the spins in the equal-spin triplet pairing of the superconductor and the easy axis $\hat z$ of the $p$-wave magnet, as shown in Fig.~\ref{fig:Gvsth}. The magnitude of the nonlocal conductivity decreases as $\th$ increases from $0$ to $\pi/2$. At $\th = \pi/2$, $G_{\mR\mL} = 0$ because the pairing is between electrons with opposite spins, and the spin component of the Cooper pair wavefunction takes the form $(|\ua\da\ra + |\da\ua\ra)$. In this case, for any incident electron on the left side, no corresponding hole of opposite spin exists on the right side. 

Interestingly, at zero bias, $G_{\mL\mL} + G_{\mR\mL} = 0$ for all $\phi$, indicating the absence of AR at the left interface. This occurs because the  zero energy modes of the triplet superconductor at the two ends hybridize and the peak in local conductivity at zero bias splits. However, this behavior changes at nonzero bias, where AR contributes to transport for all $0 < \th < \pi$, although ET is completely suppressed. We have verified that in the limit when the superconductor is very long $L\gtrsim a$, the zero bias peak appears in local conductivity. 

\section{Discussion}
In an experimental realization, the orientation of the $p$-wave magnets would be determined during fabrication and cannot be altered afterward. For different orientations, separate setups need to be fabricated. The orientation of the $p$-wave magnets can be identified through transverse spin currents, as detailed in Ref.~\cite{hedayati25}.

The chosen parameters are designed to reflect the key physical properties of a $p$-wave magnet-triplet superconductor junction. Specifically, the parameters for the $p$-wave magnet are set to ensure well-separated constant energy contours for the two spin species. Additionally, in most superconductors, the superconducting gap is typically much smaller than the chemical potential. To capture this  regime, we set the pairing strength $\Delta=0.1\mu_s$. 

It is well known that at the interface with a metal or magnet, the superconducting pairing amplitude experiences suppression~\cite{tanuma2006}. In our work, we do not explicitly account for this suppression. Our approach follows the spirit of earlier studies, such as those by Blonder, Tinkham, and Klapwijk~\cite{btk} and Sengupta et al.~\cite{sengupta2002}, which analyze tunneling conductance at metal-superconductor junctions without incorporating explicit order parameter suppression. When the suppression length is much smaller than the superconducting coherence length, the effect of pairing suppression can be neglected. Moreover, as long as the superconducting region between the two $p$-wave magnets is sufficiently long compared to the suppression length, the suppression near the junction does not qualitatively affect our results.

\section{ Summary and Outlook} 
In this work, we investigated electron transport in a junction consisting of two collinear $p$-wave magnets and a triplet superconductor. We demonstrated that CAR can be enhanced and even surpass ET by leveraging the spin-split band structure of $p$-wave magnets. Unlike conventional methods that depend on applied magnetic fields to favor CAR, our approach achieves this enhancement naturally, without requiring an external field. By arranging the $p$-wave magnets with a relative rotation of 180°, we identified a range of orientations where transverse momentum matching inhibits ET, thereby establishing a pure CAR regime. Furthermore, we showed that the chemical potential and the length of the superconductor have a significant impact on CAR, and that the angle $\theta$ between the triplet pairing spin and the easy axis of the magnets is crucial. This angle provides a means to probe the pairing symmetry in triplet superconductors.

Our findings underscore $p$-wave magnet-superconductor junctions as a promising platform for the experimental realization and control of nonlocal superconducting correlations. A crucial direction for future research is the experimental validation of these predictions in candidate materials such as CeNiAsO and Mn$_3$GaN~\cite{helle23}.

\section*{ Acknowledgments} The author thanks D. Suri for useful comments on the manuscript. The author acknowledges financial support from the ANRF (formerly SERB) Core Research Grant (CRG/2022/004311) and the University of Hyderabad Institute of Eminence PDF.

\bibliography{ref_almag}

\end{document}